# Exchange-biased quantum anomalous Hall effect

Peng Zhang[1*], Purnima P. Balakrishnan[2*], Christopher Eckberg[1,3,4,5], Peng Deng[1], Tomohiro Nozaki[6], Sukong Chong[1], Patrick Quarterman[2], Megan E. Holtz[7], Brian B. Maranville[2], Lei Pan[1], Eve Emmanouilidou[8], Ni Ni[8], Masashi Sahashi[6], Alexander Grutter[2], and Kang L. Wang[1,8,9†]

[1]*Department of Electrical and Computer Engineering, University of California Los Angeles, Los Angeles, CA 90095, USA*

[2]*NIST Center for Neutron Research, National Institute of Standards and Technology, Gaithersburg, MD 20899-6102, USA*

[3]*Fibertek Inc., Herndon, VA 20171, USA*

[4]*US Army Research Laboratory, Adelphi, MD 20783, USA*

[5]*US Army Research Laboratory, Playa Vista, CA 90094, USA*

[6]*Department of Electronic Engineering, Tohoku University, Sendai 980-8579, Japan*

[7]*Material Measurement Laboratory, National Institute of Standards and Technology (NIST), Gaithersburg, Maryland 20899, USA*

[8]*Department of Physics and Astronomy, University of California Los Angeles, Los Angeles, CA 90095, USA*

[9]*Department of Materials Science and Engineering, University of California, Los Angeles, CA 90095, USA.*

*These authors contributed equally: Peng Zhang, Purnima P. Balakrishnan

†Corresponding author: wang@ee.ucla.edu (K. L. W)

**The quantum anomalous Hall (QAH) effect is characterized by a dissipationless chiral edge state with a quantized Hall resistance at zero magnetic field. Manipulating the QAH state is of great importance in both the understanding of topological quantum physics and the implementation of dissipationless electronics. Here, we realized the QAH effect in the magnetic topological insulator Cr-doped (Bi,Sb)$_2$Te$_3$ (CBST) grown on an uncompensated antiferromagnetic insulator Al-doped Cr$_2$O$_3$. Through polarized neutron reflectometry (PNR), we find a strong exchange coupling between CBST and Al-Cr$_2$O$_3$ surface spins fixing interfacial magnetic moments perpendicular to the film plane. The interfacial coupling results in an exchange-biased QAH effect. We further demonstrate that the magnitude and sign of the exchange bias can be effectively controlled using a field training process to set the magnetization of the Al-Cr$_2$O$_3$ layer. Our work demonstrates the use of the exchange bias effect to effectively manipulate the QAH state, opening new possibilities in QAH-based spintronics.**

The quantum anomalous Hall (QAH) effect arises due to the combination of topologically nontrivial band structures and ferromagnetism-induced time-reversal symmetry breaking. Characterized by a quantized Hall resistance with vanishing longitudinal resistance at zero magnetic field, it was first observed experimentally in magnetically doped topological insulator (MTI) thin films, Cr- and V- doped (Bi,Sb)$_2$Te$_3$ [1-6], and more recently observed in the intrinsic magnetic topological material MnBi$_2$Te$_4$ [7], twisted bilayer graphene [8], and MoTe$_2$/WSe$_2$ Moiré heterostructures [9]. In addition to developing new MTIs, extensive efforts have been devoted to proximity-coupling a topological insulator (TI) or MTI material to other strongly magnetically ordered compounds in order to engineer the time-reversal symmetry breaking in topological surface states. Proximity coupling to more uniform or stronger magnetic materials has, for example, been employed to introduce ferromagnetism into otherwise nonmagnetic TIs [10-19] and to enhance the Curie temperature of MTI films [20-23]. Moreover, the proximity effect through the interface also implements exchange coupling as an additional degree of freedom through which the topological surface states can be manipulated. While the successful integration of topological and (anti)ferromagnetic layers has been demonstrated [10-27], direct manipulation of the quantized states through exchange interactions remains elusive.

Here, we report on exchange coupling between the QAH state of an MTI and an antiferromagnetic (AFM) insulator Al-doped $Cr_2O_3$. $Cr_2O_3$ is an antiferromagnetic insulator with a Néel temperature near or above room temperature and is a suitable substrate for the growth of high-quality MTI [24,26]. $Cr_2O_3$ has been reported to induce exchange bias in an adjacent MTI layer, but never simultaneously with the existence of the QAH state. On its own, $Cr_2O_3$ is of interest as a magnetoelectric material with potential applications in electrical control of either AFM domains or adjacent ferromagnetic (FM) layers through exchange coupling [28-31]. While this makes $Cr_2O_3$ promising for low energy spintronic applications, the single domain state necessary for practical applications is challenging to achieve in AFM devices. This issue may be resolved through the introduction of nonmagnetic Al dopants into the $Cr_2O_3$ thin film during the growth, which yields a ferrimagnetic-like moment also known as a parasitic magnetization. This emergent magnetization has been attributed to the preferential occupation of Al on one of the two Cr sublattices [32,33]. As illustrated in Fig. 1a, the Cr spins in pure $Cr_2O_3$ align parallel or antiparallel to [0001] forming a Néel vector along the *c*-axis, while the introduction of Al dopants breaks the symmetry of the two AFM sublattices. This unique parasitic magnetization enables deterministic control of the AFM domain structure and strengthens exchange-coupling with neighboring layers while preserving exciting properties, such as magnetoelectricity, from the parent compound.

The 50 nm thick 3.7% Al-doped $Cr_2O_3$ was deposited on α-$Al_2O_3$ (0001) substrates by magnetron sputtering, and its magnetic properties were characterized using a superconducting quantum interference device (SQUID) magnetometer. Fig. 1b shows the out-of-plane magnetization profile of the Al-doped $Cr_2O_3$ measured at multiple temperatures. Clear hysteresis loops were observed, consistent with the expected ferrimagnetic-like dopant-induced parasitic magnetization, which combined with the observation of hysteresis loops under the in-plane field with a lower saturation magnetization (supplementary Fig. S1), indicates a perpendicular magnetic anisotropy (PMA). However, a strong perpendicular exchange coupling can still be achieved in this bilayer system. The temperature evolution of the saturated magnetization in Fig. 1c indicates a critical temperature near 300 K, which is consistent with the Néel temperature ($T_N$) of 307 K [31] in pristine $Cr_2O_3$.

To fabricate the Al-$Cr_2O_3$/MTI bilayers presented in this study, Cr doped $(Bi,Sb)_2Te_3$ (CBST) layers were consequently grown on 50 nm thick Al-doped $Cr_2O_3$ using molecular beam epitaxy (MBE). CBST layers were grown with a thickness of 6 quintuple layer (QL) for transport studies, while thicker films (20 QL) were prepared for high-angle annular dark-field scanning transmission electron microscopy (HAADF-STEM) characterization and polarized neutron reflectometry (PNR) studies. Both the sputtered Al-$Cr_2O_3$ and MBE grown CBST showed good single-crystalline quality, as indicated by the streaky reflection high-energy electron diffraction (RHEED) patterns both before and after the epitaxy of CBST (Fig. 1d). An atomically sharp interface between $Al_2O_3$ and $Al-Cr_2O_3$, as well as the sharp interface to CBST layer can be seen from the HAADF-STEM in Fig. 1e and Fig. 1f, respectively.

To probe the interfacial coupling between the Al-doped $Cr_2O_3$ and CBST, we performed PNR measurements on a 20 QL CBST grown on a 50 nm Al-doped $Cr_2O_3$. PNR gives a measure of the depth-dependent magnetization of these bilayers. Specular reflectometry ($\vec{Q}$ perpendicular to the surface) is sensitive to in-plane moments only, so measurements were performed with an in-plane applied field, at a base temperature of 6 K. The summed spin-dependent reflectivity ($R^{++} + R^{--}$) and spin-dependent splitting ($R^{++} - R^{--}$) extracted from these measurements are plotted on the Fresnel scale in Fig. 2(a)-(d) along with the best-fit nuclear and magnetic profiles shown in Fig. 2(e). Overall, the PNR indicates bulk-like nuclear scattering length densities (SLDs) and extremely high-quality interfaces without the transitional growth regions recently observed in some recent studies of oxide/topological insulator heterostructures [34].

As has been previously reported in related $Cr_2O_3$/CBST structures, the interface of this system is magnetically complex. Past work [24,26] indicated antiparallel alignment of the CBST and $Cr_2O_3$ spins along with strong out-of-plane pinning of the CBST magnetization at the interface. This pinning was attributed to exchange coupling with the unusual surface reconstruction of $Cr_2O_3$ which leads to spins fixed along the [0001] direction. In this context, we note that the best-fit model indicates a region of zero in-plane magnetization at the interface between the CBST and $Al-Cr_2O_3$ at all measured fields. While the absence of any in-plane magnetization in this region could indicate the presence of a magnetic dead layer, it is also consistent with the previously reported pinning of the CBST magnetization along the film normal direction through antiparallel exchange coupling to reconstructed surface spins with giant magnetic anisotropy. Given that magnetically dead layers are rarely observed in CBST films, and a robust and repeatable exchange biasing effect in these films are later observed, we conclude that a conventional dead layer at the interface is unlikely. Rather, we propose that a strong exchange coupling between the CBST and Al-$Cr_2O_3$ surface spins pins the interfacial magnetization along the film normal where PNR is insensitive.

Further supporting this picture, the best model is one in which the magnetic moments within the CBST layer gradually cant from an out-of-plane orientation at the interface towards the in-plane direction at the top surface of the film. A schematic of the moment orientation is shown in Fig. 2f. The moments in the CBST layer, up to a maximum of 28 emu/cc (1 emu/cc = 1 kA/m) in-plane at 3 T, have a larger in-plane component canting over longer length scales as the field decreases, indicating strong exchange coupling between Cr moments within the CBST; however, this gradual moment canting is not present in the Al-$Cr_2O_3$ layer, which exhibits a uniform increase in magnetization with increasing field. This matches the magnetometry observation that Al-$Cr_2O_3$ has stronger perpendicular magnetic anisotropy than CBST. Further, the maximum in-plane magnetization at 7 K and 3 T in-plane of 20 emu/cc within the Al-$Cr_2O_3$ extracted from the PNR result matches the suppressed saturation magnetization of the in-plane hysteresis behavior measured using SQUID (supplementary Fig. S1). We conclude, therefore, that some fraction of the Al-$Cr_2O_3$ spins remain pinned along the film normal at all fields. Comparisons with alternate models, without these features, can be found in supplementary Fig. S2.

Having measured the vertical magnetization profile in Al-Cr$_2$O$_3$/CBST bilayers, we carried out transport measurements to show the exchange-biased QAH effect in these heterostructures. The films were patterned into Hall-bar devices with a geometry of 1 mm by 0.5 mm, as previously shown in Fig. 1g, using a hard mask and dry etching method. The samples were cooled from 300 K to 50 mK with a +1 T perpendicular magnetic field applied. Fig. 3a shows a vanishing longitudinal resistance $\rho_{xx}$ and quantized Hall resistance $\rho_{xy}$ at cryogenic temperature, confirming these films do indeed display the typical signatures of the QAH effect. For further confirmation, conductivity tensor components were computed using the relations $\sigma_{xx} = \rho_{xx}/(\rho_{xx}^2 + \rho_{xy}^2)$ and $\sigma_{xy} = \rho_{xy}/(\rho_{xx}^2 + \rho_{xy}^2)$, indicating perfect quantization (Fig. 3b). The flow diagram in Fig. 3c shows the evolution of $\sigma_{xx}$ versus $\sigma_{xy}$ upon a magnetic field sweep between $\pm 0.2$ T at the base temperature of 50 mK. It showed the transition between the two quantized Chern insulator states with Chern number $c = \pm 1$, $(\sigma_{xy}, \sigma_{xx}) = (c, 0)$, and an intermediate trivial insulator state, $(\sigma_{xy}, \sigma_{xx}) \to (0,0)$, which is observed in QAH insulators with thickness $t \leq 6$ QL due to a competition between the magnetic exchange gap and surface hybridization gap [3,35,36]. The sample was then subjected to a small magnetic field sweep between $\pm 0.2$ T. The field sweep range was larger than the coercivity of CBST but smaller than that of Al-Cr$_2$O$_3$, such that the magnetization of the latter remained intact during the sweep. The sample showed an exchange-biased quantum anomalous Hall effect as evidenced by a shift of the coercivity field in both $\rho_{xx}$ (Fig. 3d) and $\rho_{xy}$ (Fig. 3e) under different Al-Cr$_2$O$_3$ magnetization initialization status. The exchange bias has a magnitude of around $2\mu_0 H_{eb} \approx 16$ mT, and a slightly asymmetric behavior of $\rho_{xx}$, which fully reverses, could be due to different spin texture at the Al-Cr$_2$O$_3$ and CBST interface under different field training conditions. Remarkably, the exchange bias was observed simultaneously with the QAH effect, showing that the presence of the exchange bias from the Al-Cr$_2$O$_3$ layer could effectively manipulate the QAH state in CBST.

We would like to emphasize that the exchange-biased QAH effect can be controlled by a novel field-training (FT) approach, rather than requiring the traditional field cooling (FC) method. Field cooling is usually used to align the Néel vector of AFM materials, by applying a magnetic field and cooling through the Néel temperature. In Al-Cr$_2$O$_3$, the introduction of Al dopants breaks the symmetry of the two AFM sublattices and produces a nonzero net magnetization as indicated in the SQUID and PNR experiments. While an external magnetic field cannot couple to a perfectly compensated AFM order parameter, it can be used to switch the ferrimagnetic-like magnetic order in Al-Cr$_2$O$_3$. The use of Al-Cr$_2$O$_3$ thus has a significant advantage over its undoped counterpart Cr$_2$O$_3$ as the sign of any exchange bias may be changed through the application of an external magnetic field alone, without requiring the system be raised above its Néel temperature $T_N$.

To demonstrate this advantage, we show that the magnetism in Al-Cr$_2$O$_3$, and consequently the sign of the interfacial exchange bias, may be initialized using both field training and traditional field cooling methods. For field cooling, the sample was subjected to $\pm 3$ T fields while cooling down from above $T_N$ = 330 K to 2 K. For field training, the sample was cooled down to 2 K without applying field and then was subjected to $\pm 3$ T and $\pm 9$ T fields. In all Al-Cr$_2$O$_3$ initialization scenarios, small magnetic field sweeps between ±0.2 T were then performed at 2 K. As shown in Fig. 4a, after +3 T field cooling, +3 T field training, or +9 T field training, the device displays the same behavior; most notably, a pronounced exchange bias when compared to the similar field cooling or field training process using a negative field, indicating that field training and field cooling are equally effective methods for establishing exchange bias in this system.

The exchange bias was shown to be directly related to Al-$Cr_2O_3$ magnetization from the dependence of exchange bias on the externally applied training field. Before each small field sweep (±0.2 T), a negative exchange bias was set using a +3 T field, and the sample was subsequently subjected to a training field of opposing sign $-\mu_0 H_{train}$ (Fig. 4b). The magnitude of this training field was varied between 0.2 T and 3 T, resulting in a gradual switching of Al-$Cr_2O_3$ magnetization. Shown in Fig. 4b is that the anomalous Hall loop of CBST gradually shifts to the positive direction and finally saturates with increasing training field, indicating a sign change in the exchange bias. In this measurement series, zero exchange bias occurs with a training field of 0.7 T, whereas the value of the exchange bias saturates with a training field of approximately 1 T. These values correlate approximately with the coercivity and switching fields of Al-$Cr_2O_3$ obtained from the SQUID data in Fig. 1b.

The same measurement was performed at different temperatures, and the exchange bias as a function of training field magnitude was obtained and shown in the color plot in Fig. 4c. The inset schematically illustrated the different signs of exchange bias under different training fields due to the antiparallel alignment of the CBST and Al-$Cr_2O_3$ surface spins. The temperature dependence of $|H_{eb}|$ is summarized in Fig. 4d, where the exchange bias magnitude decreases with increasing temperature and vanishes when the temperature reaches above the Curie temperature (≈30 K) of CBST thin film. This observation, combined with the absence of magnetic proximity effects in an undoped BST and Al-$Cr_2O_3$ heterostructure (supplementary Fig. S4), indicates the exchange bias is most likely due to the exchange coupling of Cr atoms between Al-$Cr_2O_3$ and CBST.

**To summarize**, we demonstrate high quality MBE growth of CBST on an uncompensated AFM insulator substrate (Al-$Cr_2O_3$). PNR measurements reveal a vanishing in-plane magnetization at the Al-$Cr_2O_3$ and CBST interface, which, when taken together with the transport measurements, indicates a strong exchange coupling between the two magnetic layers pinning the moments at the interface to point out-of-plane. Interestingly, while earlier studies on undoped $Cr_2O_3$/MTI heterostructures reported that exchange bias and the QAHE were never observed simultaneously, the strong exchange coupling in Al-$Cr_2O_3$/CBST results in the ability to exchange bias the quantum anomalous Hall effect, which could facilitate the development of functional topological spintronic devices when combined with the magnetoelectric effect of Al-$Cr_2O_3$ [32,33]. We further demonstrate that the magnitude and sign of the exchange bias can be effectively controlled using a field training process to set the magnetization of the Al-$Cr_2O_3$ layer, so that it is possible to directly manipulate the exchange bias without modifying the temperature. Thus, we can magnetically tune both the magnetic and topological properties of the MTI. This work successfully demonstrates the manipulation of QAH states by an adjacent magnetic layer through exchange coupling, which provides an additional degree of freedom through which the topological surface states can be manipulated. Our findings highlight the rich tuning possibilities in topological and AFM heterostructures, and this approach can also be used to understand and manipulate other emerging topological quantum phases such as axion insulator states [37-39] and high-Chern number QAH states [40].

**Methods**

The single crystal Al-doped $Cr_2O_3$ (50nm) thin films were grown on $Al_2O_3$ (0001) substrate using the reactive sputtering method in an $Ar + O_2$ atmosphere at a substrate temperature of 773 K with a $Al_5Cr_{95}$ alloy target [32,33]. The Al-doped $Cr_2O_3$ films were then transferred through air, and high-quality Cr-doped $Bi_xSb_{2-x}Te_3$ thin films were grown using a Perkin Elmer MBE system in ultrahigh vacuum. High-purity Bi (99.9999%), Te (99.9999%), Cr (99.99%) and Sb (99.999%) were evaporated by conventional effusion cells and cracker cells. During the growth, the substrate was maintained at a temperature of 200 °C. The growth was monitored by an in-situ reflection high-energy electron diffraction (RHEED. The SQUID characterization is done in a Quantum Design Magnetic Properties Measurement System (MPMS3). SQUID data are presented with linear background associated with the substrate's diamagnetic signature subtracted. Grown films were patterned into Hall bars for the transport study using a hard mask with dry etching methods. Low-frequency four-probe magnetoresistance measurements and Hall measurements were conducted in a Quantum Design Physical Properties Measurement System (PPMS) with a dilution refrigerator insert. Polarized neutron reflectometry (PNR) measurements were performed with the Polarized Beam Reflectometer (PBR) instrument at the NIST Center for Neutron Research. Samples were field-cooled, in an applied field of 3 T in-plane, to a temperature of 6 K; spin-up and spin-down non-spin-flip reflectivities (↑↑ and ↓↓) were then measured in a specular configuration at in-plane fields of 3 T, 250 mT, and 110 mT. Data was reduced using the Reductus software package [41], and then fit simultaneously using Refl1d [42] to model the nuclear and magnetic scattering length densities (SLD) as a function of depth within the sample; further details can be found in the supplemental section S2.

**Acknowledgements**

This work was supported by the NSF under Grants No. 1936383 and No. 2040737, the U.S. Army Research Office MURI program under Grants No. W911NF-20-2-0166 and No. W911NF-16-1-0472. C.E. is an employee of Fibertek, Inc. and performs in support of Contract No. W15P7T19D0038, Delivery Order W911-QX-20-F-0023. The views expressed are those of the authors and do not reflect the official policy or position of the Department of Defense or the US government. The identification of any commercial product or tradename does not imply endorsement or recommendation by Fibertek Inc. T.N. and M.S. were funded by the ImPACT Program of Council for Science, Technology and Innovation (Cabinet Office, Japan Government). E.E and N.N. were supported by the U.S. Department of Energy (DOE), Office of Science, Office of Basic Energy Sciences under Award No. DE-SC0021117. The research was performed in part at the National Institute of Standards and Technology (NIST). Any mention of specific trade names and commercial products is for information and experimental reproducibility only; it does not imply recommendation or endorsement by NIST.

**Author Contributions**

P.Z. and K.L.W. conceived and designed the research. P.Z., L.P., T.N. and M.S performed the sample growth. M.E.H. performed the TEM characterization. C. E. performed the SQUID measurements. S.C. fabricated the device. P.Z. and P.D. performed transport measurements with the help of E.E. and N.N. P.Z. analysed the transport data. P.P.B, P.Q., B.B.M. and A.G performed the PNR measurement and data analysis. P.Z., P.P.B., A.G and K.L.W. wrote the manuscript with the inputs from all the other co-authors.

**Data availability**

The data of the study are available from the corresponding author upon reasonable request.

**Competing interests**

The authors declare no competing financial interests.

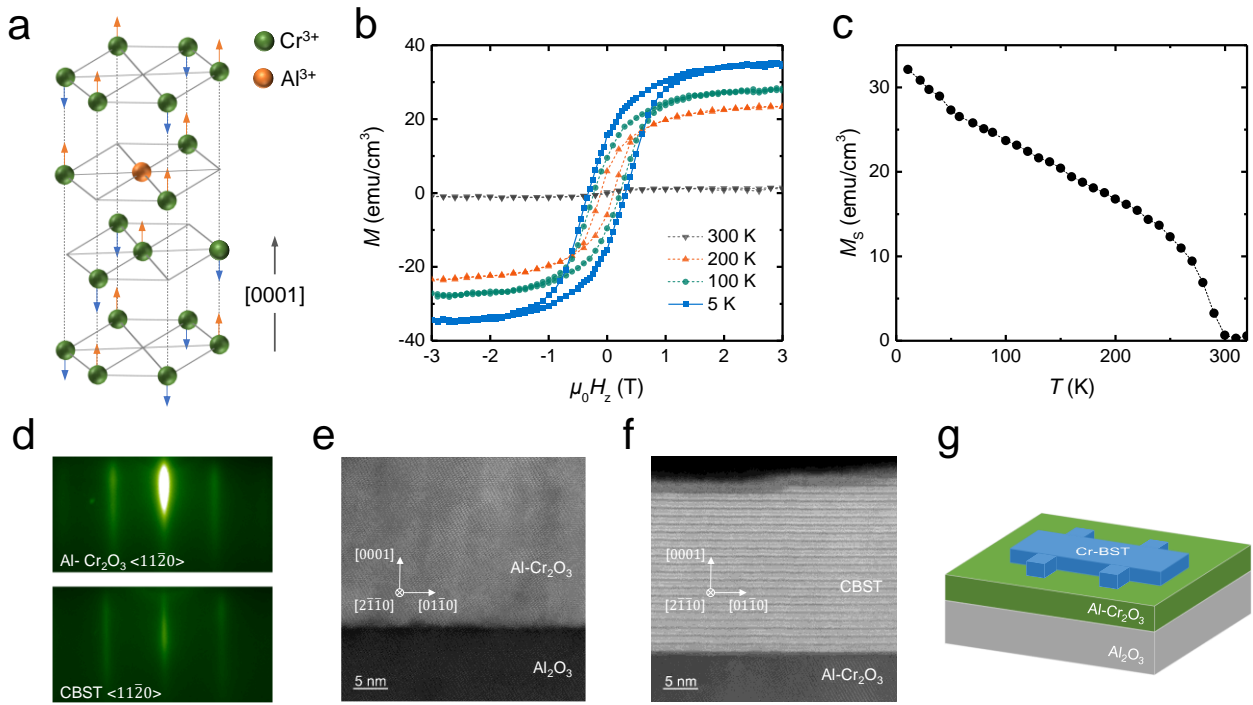

**Figure 1. High quality Cr doped BST grown by molecular beam epitaxy on Al-doped $Cr_2O_3$ (0001). a.** Atomic and magnetic structure of Al-doped $Cr_2O_3$. Oxygen atoms are omitted for simplicity. $Al^{3+}$ substitution of $Cr^{3+}$ change the AFM $Cr_2O_3$ into a ferrimagnetic-like magnetic structure. **b.** M-$H_z$ profile of 50 nm Al-$Cr_2O_3$ substrate showing hysteresis loops below 300 K. A linear, diamagnetic background associated with the $Al_2O_3$ substrate has been subtracted from all presented curves. **c.** Temperature dependence of saturated magnetic moment of 50 nm Al-$Cr_2O_3$ substrate measured with an out-of-plane filed of +2 T. **d.** RHEED pattern of the Al-$Cr_2O_3$ substrates and MBE grown CBST on top along <11-20> direction. **e.** HAADF-STEM of the $Al_2O_3$/Al-$Cr_2O_3$ interface. **f.** HAADF-STEM of the Al-$Cr_2O_3$/CBST interface. **g.** Hall-bar structure (1 mm by 0.5 mm) used for transport measurement.

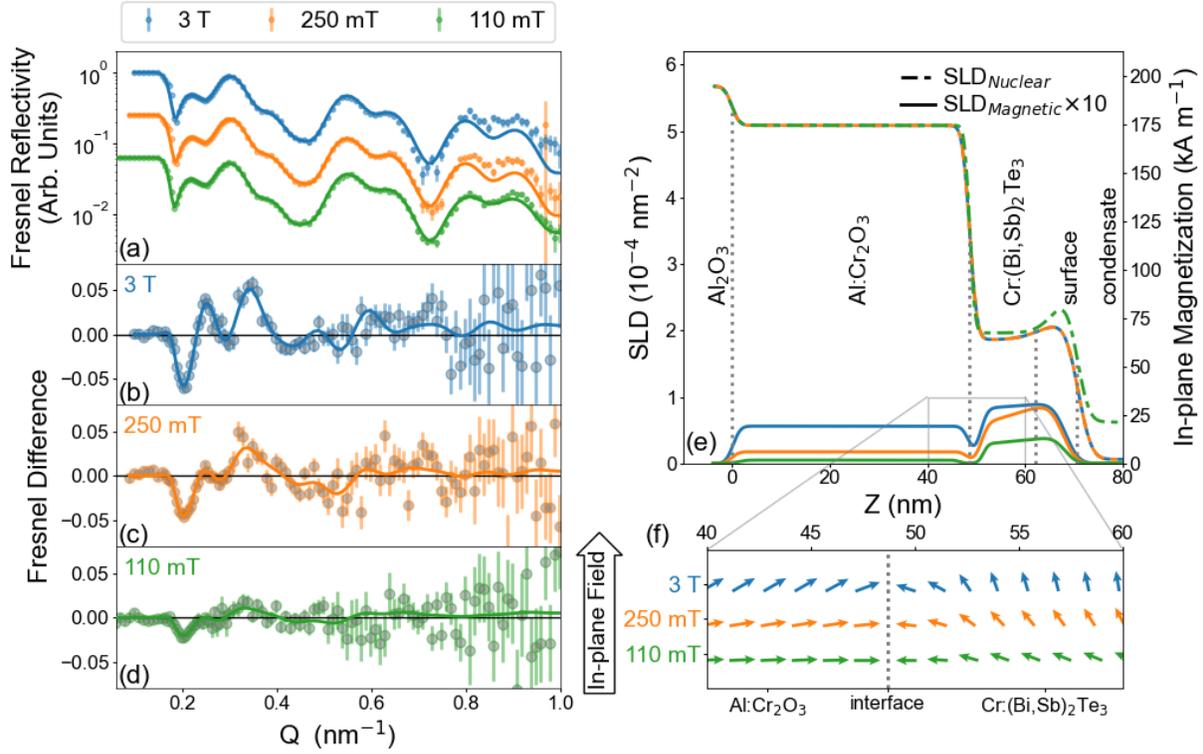

**Figure 2. Interfacial coupling between Al-Cr$_2$O$_3$ and CBST layers probed by PNR. a.** Fresnel reflectivity $\left(\frac{R^{++}+R^{--}}{R(Al_2O_3)}\right)$ scaled to be offset and **b-d.** difference $\left(\frac{R^{++}-R^{--}}{R(Al_2O_3)}\right)$ for an Al-Cr$_2$O$_3$/CBST bilayer measured by polarized neutron reflectometry with applied in-plane magnetic fields of b. 3 T (blue), c. 250 mT (orange), and d. 110 mT (green). Solid lines show the best fits, which are generated by **e.** nuclear and in-plane magnetic depth-profile models at each field. Magnetic scattering length density (SLD) has been scaled by a factor of 10 for clarity but corresponds to the magnetization as plotted. Differences in the nuclear SLD profiles at the surface between subsequent measurements are likely due to sample aging, as discussed in Fig. S3. **f.** Possible orientation and relative magnitude of the net magnetization near the interface, deduced from (e), the in-plane magnetization, and magnetometry results. This schematic assumes that the Al-Cr$_2$O$_3$ sublattices rotate together and does not illustrate the effect of domains. All error bars represent ±1 standard deviation.

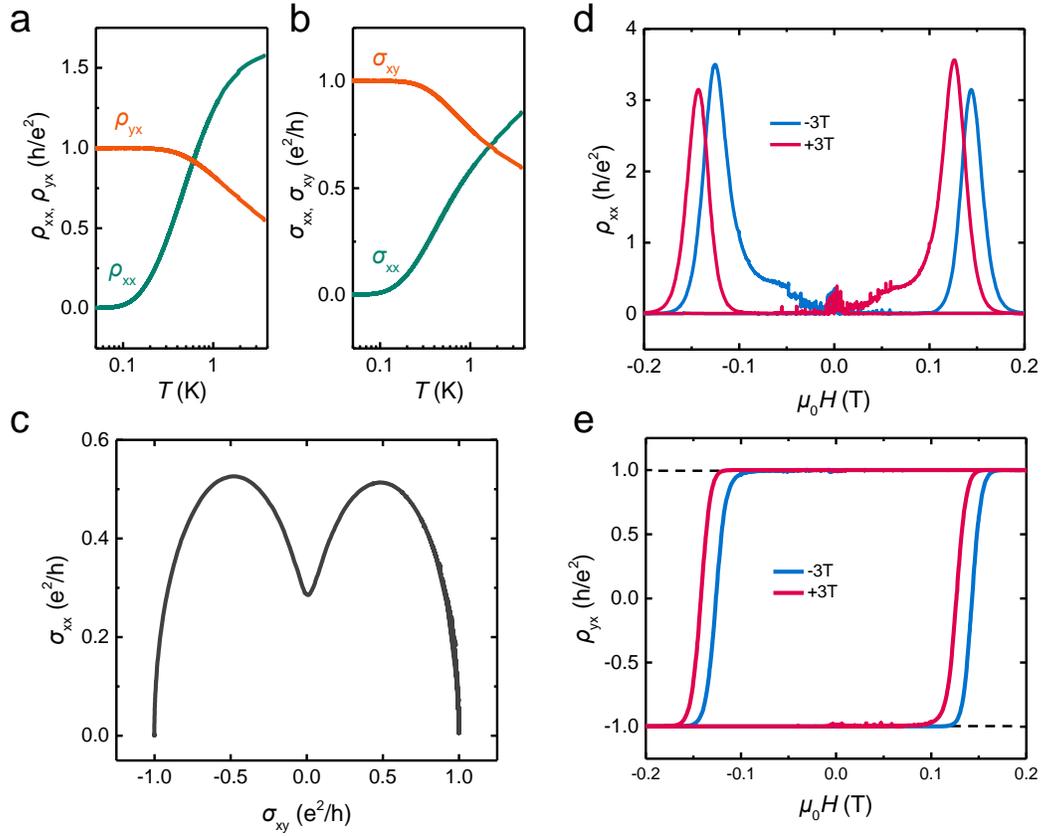

**Figure 3. Observation of exchange-biased quantum anomalous Hall effect in Al-Cr$_2$O$_3$/CBST heterostructure. a.** Temperature dependences of longitudinal resistivity $\rho_{xx}$ and Hall resistivity $\rho_{yx}$ under 1 T field, where $\rho_{xx}$ vanishes and $\rho_{yx}$ approaches the quantized value of h/e$^2$ with decreasing temperature. **b.** Temperature dependences of $\sigma_{xx}$ and $\sigma_{xy}$ converted from $\rho_{xx}$ and $\rho_{yx}$. **c.** Evolution of the $\sigma_{xx}$ versus $\sigma_{xy}$ when scanning the magnetic field at 50mK. **d** and **e.** Magnetic field dependences of $\rho_{xx}$ and $\rho_{yx}$ after training the sample using -3 T and +3 T to polarize the Al-Cr$_2$O$_3$ magnetization. The magnetic field was swept between ±0.2 T and the data was taken at 50 mK.

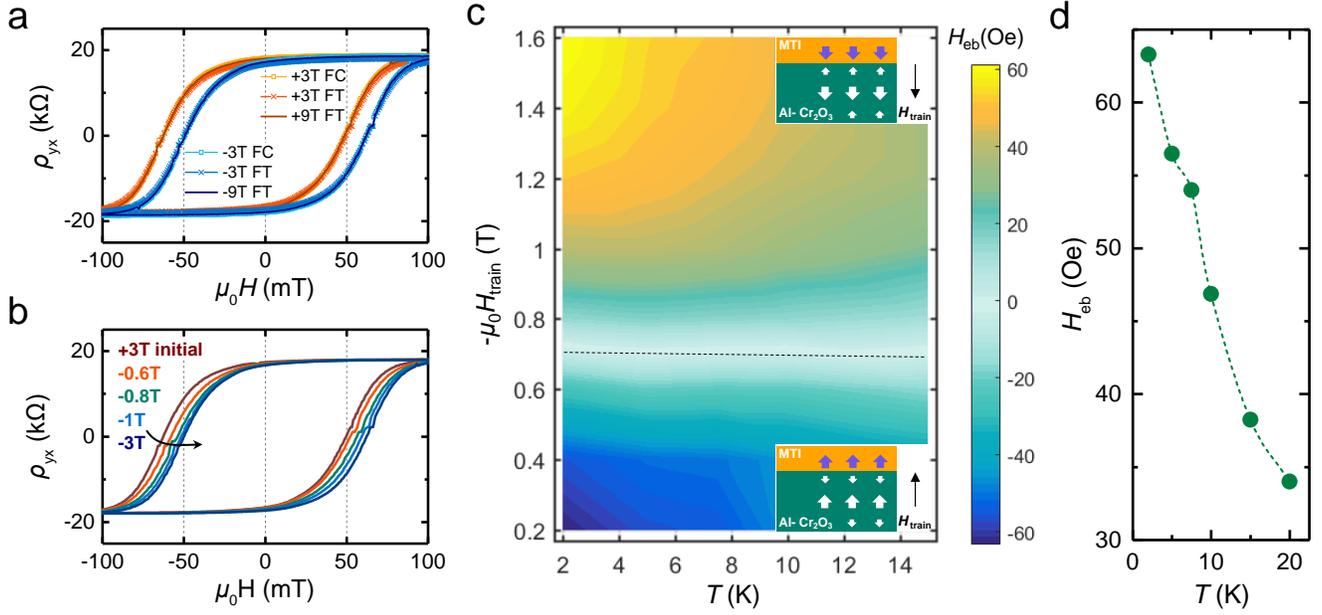

**Figure 4. Exchange bias manipulated by field-training of Al-Cr$_2$O$_3$. a.** Small field sweep (±0.2 T) of $\rho_{yx}$ at $T$= 2 K, after treating the sample with ±3 T, ±9 T field training (FT) and ±3 T field-cooling (FC) processes. **b.** Small field sweep (±0.2 T) of $\rho_{yx}$ at $T$= 2 K, after sample treated with various training field. +3 T field is applied to initialize the magnetization of Al-Cr$_2$O$_3$ before each field-training and loop scan. **c.** Mapping of the training field dependent exchange bias $H_{eb}$ at different temperatures. The gradual switching of $H_{eb}$ with increasing training field corresponds to the switching of the Al-Cr$_2$O$_3$ magnetization. **d.** Summarized temperature dependence of $H_{eb}$. The same device that has been described in Fig. 3 is studied for this measurement.